\documentclass[12pt]{iopart}

\usepackage{graphicx}
\begin{document}

\title[$\phi$-mesons in high energy nuclear collisions]{Probe the QCD phase diagram with $\phi$-mesons in high energy nuclear collisions}

\author{B. Mohanty$^{a}$ and N. Xu$^{b}$}

\address{$^{a}$ Variable Energy Cyclotron Centre, 1/AF Bidhan Nagar, Kolkata - 700064, India
         and $^{b}$ Lawrence Berkeley National Laboratory, 1 Cyclotron Road, California - 94720, USA}
\ead{$^{a}$bmohanty@veccal.ernet.in}
\begin{abstract}

High-energy nuclear collision provide a unique tool to study the strongly interacting 
medium. Recent results from the Relativistic Heavy Ion Collider (RHIC) on $\phi$-meson
production has revealed the formation of a dense partonic medium. The medium constituents
are found to exhibit collective behaviour initiated due to partonic interactions in the 
medium. We present a brief review of the recent results on $\phi$ production in heavy-ion 
collisions at RHIC. One crucial question is where, in the phase diagram, does the 
transition happen for the matter changing from hadronic to partonic degrees of freedom. 
We discuss how  $\phi$-meson elliptic flow in heavy-ion collisions can be used for 
the search of the QCD phase boundary. 
\end{abstract}


\section{Introduction}
$\phi$-mesons were first observed in bubble chamber experiments at Brookhaven
National Laboratory (BNL) in the year 1962 through the reaction 
$K^{-} + p \longrightarrow \Lambda + K + \bar{K}$. Results 
on $\phi$-meson production at the Relativistic Heavy Ion Collider (RHIC) at BNL 
has so far provided deep insight on the nature  of the medium formed in heavy
ion collisions at high energies~\cite{star,phenix}. The $\phi$-meson has several interesting features
and each of these features are capable of providing valuable information about the
medium properties in high energy heavy-ion collisions. Some of these features and possible
information they provide are tabulated in Table~\ref{table1}. Hence $\phi$-mesons can be termed as
a golden tool (term borrowed from golden ratio used in Mathematics and Art~\cite{goldenratio}) to 
address various aspects of heavy-ion collisions.

In this paper we first review the existing results on $\phi$ production at RHIC.
These will be presented with a motivation of establishing observations regarding 
$\phi$ production (through rapidity, transverse momentum and azimuthal distribution) as a 
clear indication of dense partonic medium formation at RHIC energies. The results 
presented are mostly from the hadronic decay mode $\phi$ $\rightarrow$ $K^{+}$$K^{-}$ 
in Au+Au and Cu+Cu collisions at midrapidity~\cite{star,phenix}. We will first discuss three physics
issues: (a) strangess enhancement, (b) partonic recombination 
and hints of thermalization at RHIC and (c) partonic collectivity using $\phi$ production.  
These will provide the basis for our proposal of using the $\phi$-meson elliptic flow ($v_{2}$) 
as a probe of QCD phase boundary, which will be discussed next in the paper. Finally
we will conclude by sumarizing the observation on $\phi$ production at RHIC and 
possibilities it offers for the future high energy heavy-ion collision program.

\begin{table}[hbt]
\caption{\label{table1}Various features of $\phi$-meson and possible information 
they may provide about the medium formed in high energheavy-ion collisions}
\begin{tabular}{@{}|c|c|}
\br
Feature&Information \\
\mr
Quark content : $s\bar{s}$& Strangeness enhancement; net strangeness\\
                          & $\Delta S = 0$ Canonical suppression not applicable.\\
\hline
Meson, Mass = 1.019 GeV/$c^{2}$ & Differentiates Mass and Number of constitutent\\ 
                                & quark (NQ) effects; Mass $\sim$  baryons 
                                  ($p$ and $\Lambda$)\\
Width = 4.43 MeV/$c^{2}$ & Experimentally clean signal; Change in width \\
                         & reflects medium effects\\
\hline
Decay modes & Both Hadronic and Leptonic; Leptons do not \\
            & interact strongly with the medium.\\
Primodial fraction & $\sim$ 100\%;not affected by resonance decays\\
\hline
Small interaction with nucleons & Early freeze-out; Information of partonic stage\\
\hline
Life time $\sim$ 45 fm/$c$ & With $K^{*}$ (lifetime $\sim$ 4 fm/$c$) ideal to understand\\
                           &  rescattering and regeneration effects for resonances\\
\br
\end{tabular}
\end{table}
\normalsize

\section{$\phi$-meson yields - strangeness enhancement}
The large abundances of $s$ and $\bar{s}$ quarks in the Quark Gluon Plasma (QGP), may lead 
to a dramatic increase in the production of $\phi$-mesons and other strange hadrons 
relative to non-QGP $p$+$p$ collisions~\cite{seqgp}.
Alternative ideas of canonical suppression of strangeness in small systems as a source
of strangeness enhancement in high energy heavy-ion collisions have been proposed for 
other strange hadrons (e.g $\Lambda$, $\Xi$ and $\Omega$)~\cite{canonical}.
According to these models, strangeness enhancement in nucleus-nucleus 
collisions, relative to $p$+$p$ collisions, should increase with the strange quark content of 
the hadrons. This enhancement is predicted to decrease with increasing beam 
energy~\cite{canonical}. So far, discriminating between the two scenarios 
(strange hadron enhancement being due to dense partonic  medium formed in heavy-ion 
collisions or due to canonical supression of their production in $p$+$p$ collisions) using 
the available experimental data has been, to some extent,  ambiguous. $\phi$-mesons  
due to its zero net strangeness is not subjected to Canonical suppression effects. 
Enhancement of $\phi (s\bar{s})$ production,in Cu+Cu and Au+Au relative to $p$+$p$ 
collisions would clearly indicate the formation of a dense partonic medium in these 
collisions. This would then rule out canonical suppression effects being the most 
likely cause for the observed enhancement in other strange hadrons in high energy 
heavy-ion collisions.

\begin{figure}
\begin{center}
\includegraphics[scale=0.3]{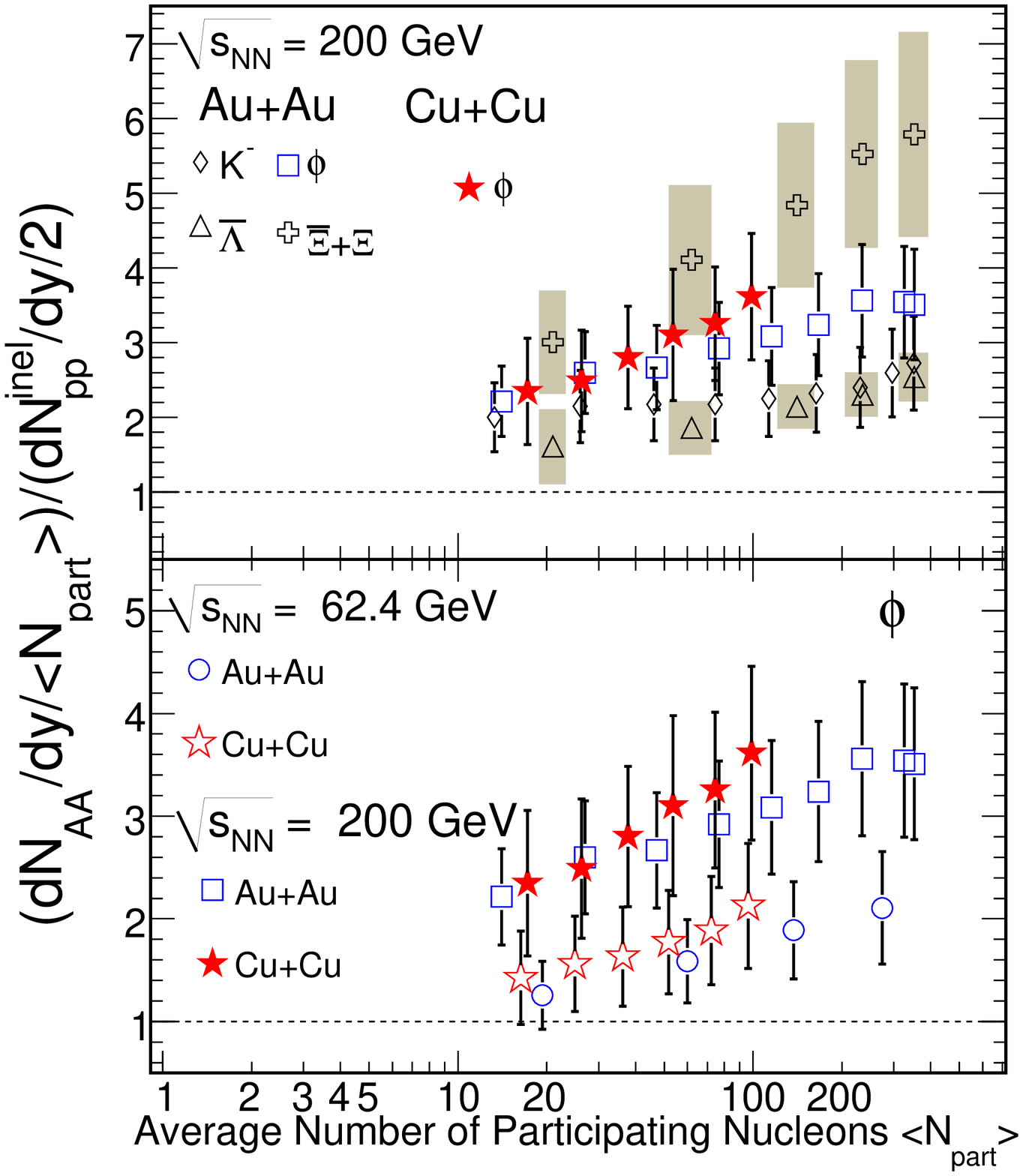}
  \includegraphics[scale=0.35]{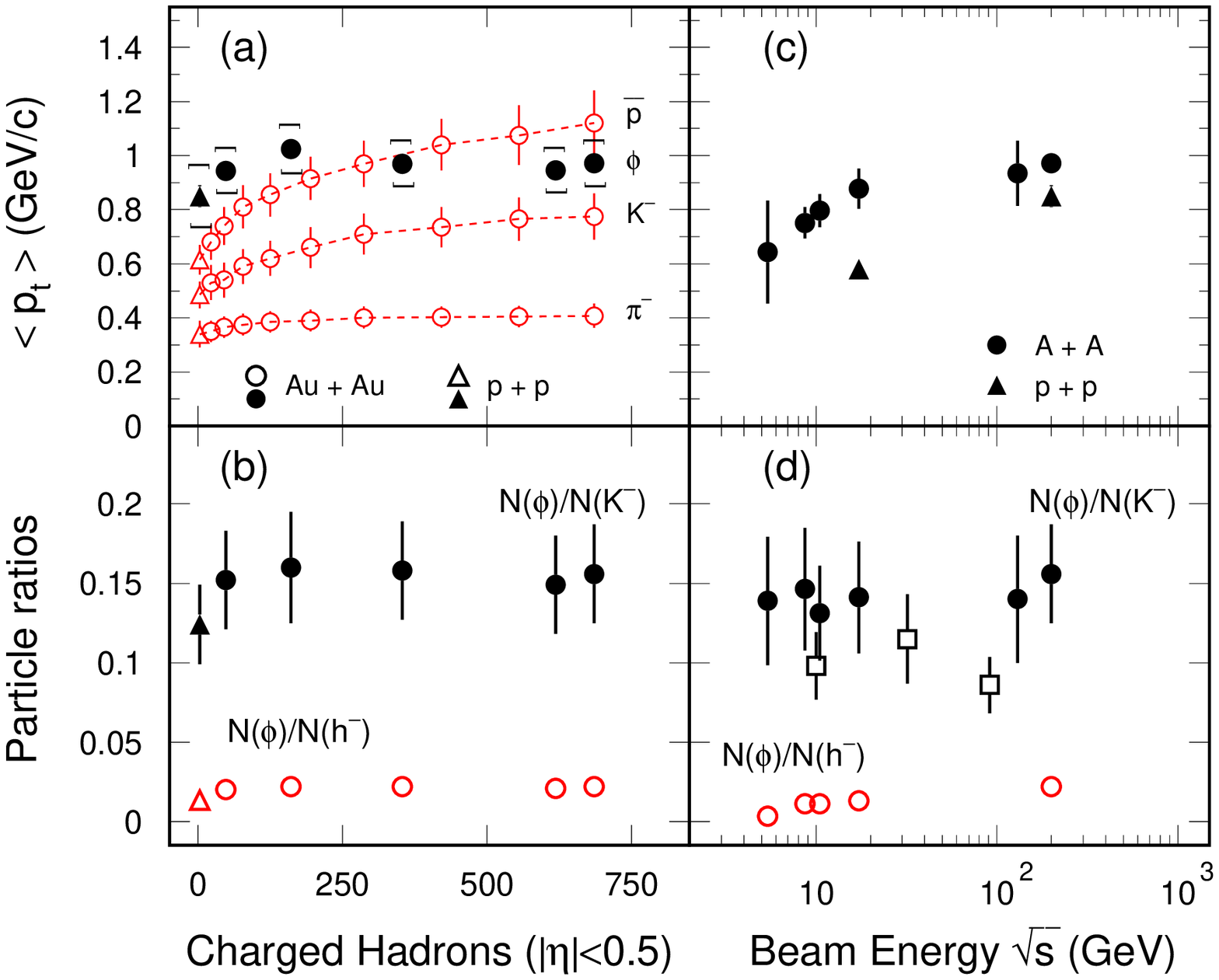}
  \includegraphics[scale=0.25]{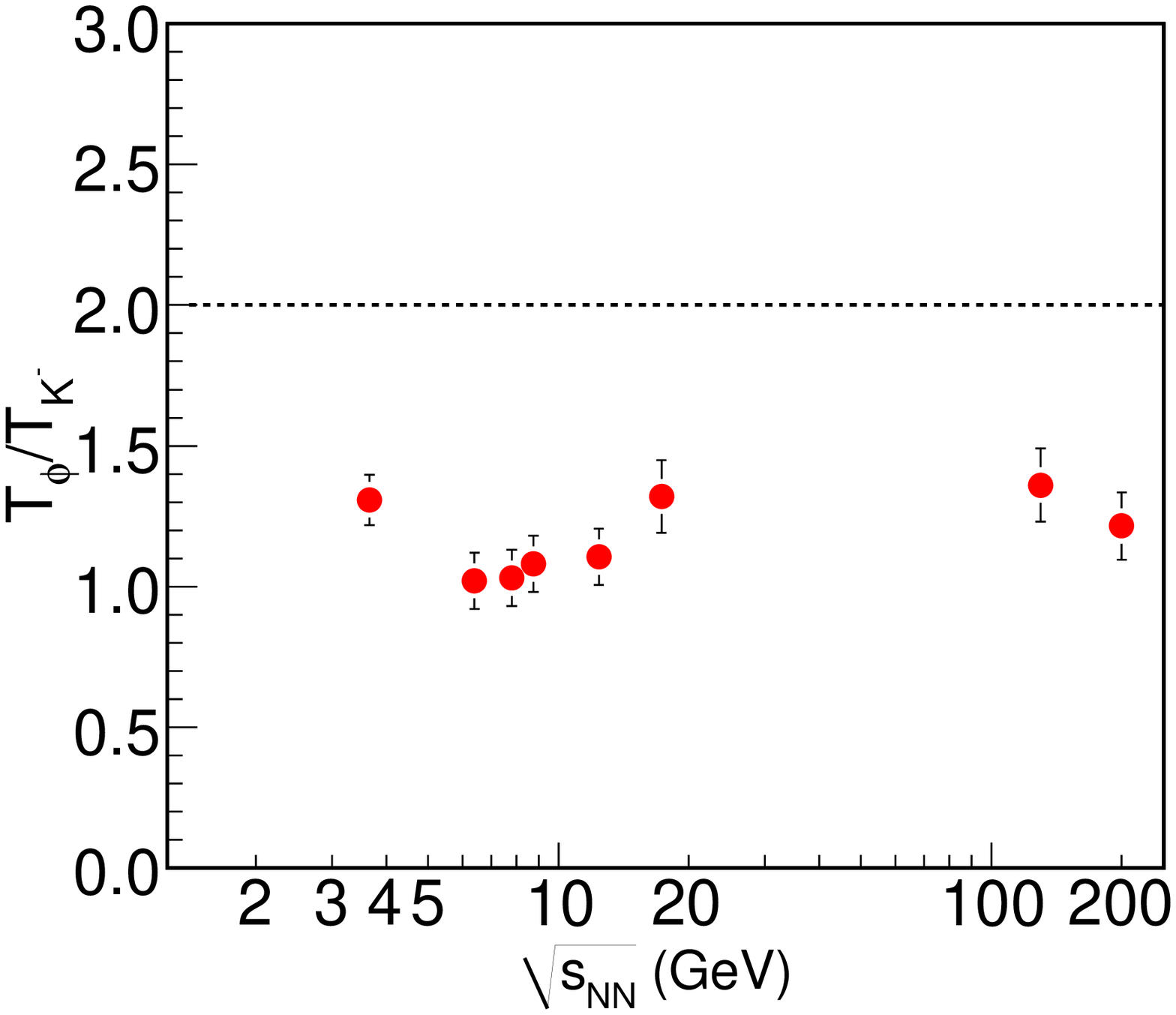}
  \includegraphics[scale=0.25]{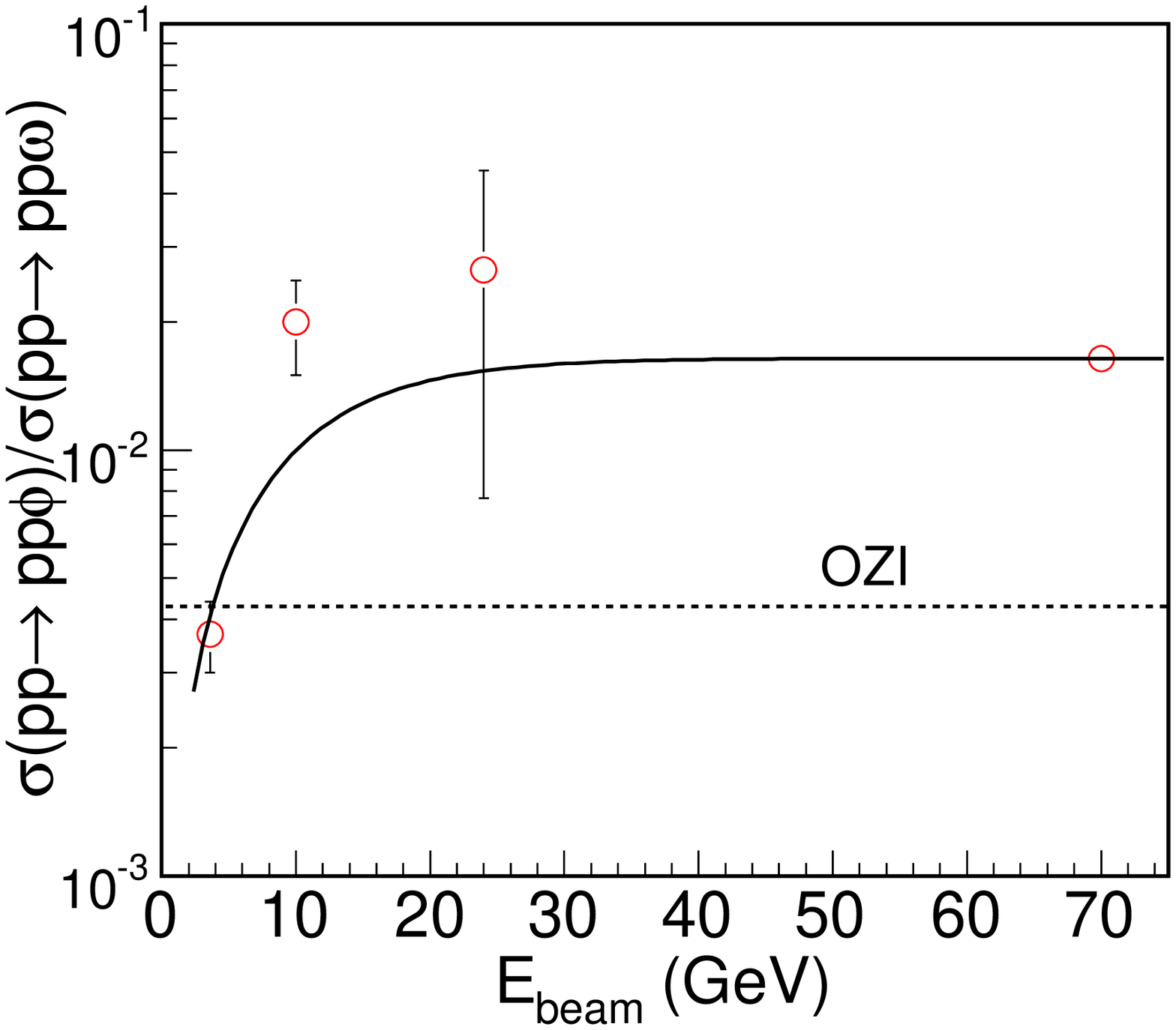}
\caption{(color online) Top Left : Upper panel: The ratio of the yields of $K^{-}$, $\phi$, $\bar{\Lambda}$ 
and  $\Xi+\bar{\Xi}$ normalized to $\langle N_{\mathrm {part}} \rangle$ in nucleus-nucleus 
collisions to corresponding yields in inelastic $p$+$p$ collisions as a function 
of $\langle N_{\mathrm {part}} \rangle$ at 200 GeV. Lower panel: Same as above 
for $\phi$-mesons in Cu+Cu collisions at 200 and 62.4 GeV. 
Top Right : Average transverse momentum ($\langle p_{T} \rangle$) for produced hadrons
in Au+Au and p+p collisions as function of collision centrality at 200 GeV and as a function
of beam energy. $N(\phi)/N(K^{-})$ ratio and $N(\phi)/N(h^{-})$ as a
function of collision energy and collision centrality in high energy collisions.
Bottom Left : Ratio of slope of transverse momentum distribution of $\phi$-mesons
to $K^{-}$ as a function of beam energy in heavy-ion collisions.
Bottom Right : $N(\phi)/N(\omega)$ ratio as function of beam energy in $p$+$p$ collisions.
The line shows the expectation due to OZI rule~\cite{ozi}.
}
\label{fig6}
\end{center}
\end{figure}

Figure~\ref{fig6} shows the ratio of strange hadron production normalized to 
$\langle N_{\mathrm {part}} \rangle$ in nucleus-nucleus collisions relative to corresponding
results from $p$+$p$ collisions at 200 GeV. The results are plotted as a function 
of $\langle N_{\mathrm {part}} \rangle$. $K^{-}$, $\bar{\Lambda}$ and  
$\Xi+\bar{\Xi}$~\cite{star} are seen to show an enhancement (value $>$ 1) that increases 
with the number of strange valence quarks. However, the enhancement of $\phi$-meson production 
from Cu+Cu and Au+Au collisions shows a deviation in ordering in terms of the number of strange 
constituent quarks~\cite{becatini}. 
Enhancement is larger than for $K^{-}$ and $\bar{\Lambda}$, at the
same time being smaller than in case of $\Xi+\bar{\Xi}$. 
So the  $\phi$-mesons do not follow the strange quark ordering as expected in the canonical picture 
for the production of other strange hadrons. The observed enhancement in $\phi$-meson production being 
related to medium density is further supported by the energy dependence shown in the lower panel and
top-left plot of 
 Fig.~\ref{fig6} . The $\phi$-meson production relative to $p$+$p$ collisions is larger at higher 
beam energy, a trend opposite to that predicted in canonical models for other strange hadrons. 

To establish that the observed enhancement of $\phi$-meson production is a clear indication 
of the formation of a dense partonic medium in Au+Au collisions at 200 GeV we need to establish 
the following : (a) $\phi$ production is not from K$\bar{K}$ coalesence (as Kaons could be 
subjected to canonical suppression effects) and (b) $\phi$ production is not OZI suppressed 
in p+p collisions. There are several experimental observations which shows that at 
RHIC energies $\phi$ production is not dominantly from K$\bar{K}$ coalesence. These 
include: (i) $N(\phi)/N(K^{-})$ ratio is observed to be independent of colliding beam energy and
 at a given beam energy independent of collision centrality (Fig.~\ref{fig6} top right)~\cite{star}.
Naive expectation from  K$\bar{K}$ coalesence models is a linear increase of the $N(\phi)/N(K^{-})$ ratio
with increase in collision centrality~\cite{rqmd}; (ii) If $\phi$ production is dominantly from 
K$\bar{K}$ coalesence, 
one expects the ratio of inverse slope of transverse momentum distribution of $\phi$ to those 
of $K^{-}$ $\sim$ 2. This is not observed from the available experimental data (Fig~\ref{fig6} 
bottom left); (iii) If $\phi$ production is dominantly from K$\bar{K}$ coalesence,
one expects the width of the rapidity distribution of $\phi$-mesons to be related to those 
for charged Kaons as
1/$\sigma_{\phi}^{2}$ = 1/$\sigma_{K^{-}}^{2}$ + 1/$\sigma_{K^{+}}^{2}$. 
Measurements at SPS energies show a clear deviation of the data from the above expectation~\cite{NA49}; 
(iv) Finally, if $\phi$ production is dominantly from K$\bar{K}$  coalesence it 
would be reflected in elliptic flow ($v_{2}$) measurements. We observe at intermediate $p_{T}$, 
the $v_{2}$ of $\phi$-mesons and Kaons are comparable (discussed later in Fig.~\ref{fig:V2})~\cite{star,phenix}.
Now about OZI suppression of $\phi$ roduction in $p$+$p$ collisions. These effects are 
experimentally observed by measuring the ratio of $\phi$ to $\omega$ production. This ratio 
takes up a value of 4.2 $\times$ $10^{-3}$ for validity of OZI rules. Violations
of OZI rule~\cite{ozi} has been observed in $p$+$p$ collisions at higher energies as shown in 
Fig.~\ref{fig6} (bottom right)~\cite{pp_ozi}. Due to the high gluon density created in 
high energy collisions, the OZI suppression does not play a role in $p$+$p$ collisions
at RHIC. So it is very unlikely that  $\phi$ enhancement observed 
in Au+Au collisions is due to OZI suppression of its production
in $p$+$p$ collisions. Thus with both the physics points discussed in (a) and (b) above established,
the observed enhancement of $\phi$-meson production 
then is a clear indication that the strangeness enhancement in Au+Au collisions at 200 GeV is
due to the formation of a dense partonic medium.

\section{$\phi$-meson production - quark recombination and thermalization}

\begin{figure}[!hbt]
\vspace{-0.45cm}
  \center
  \includegraphics[scale=0.25]{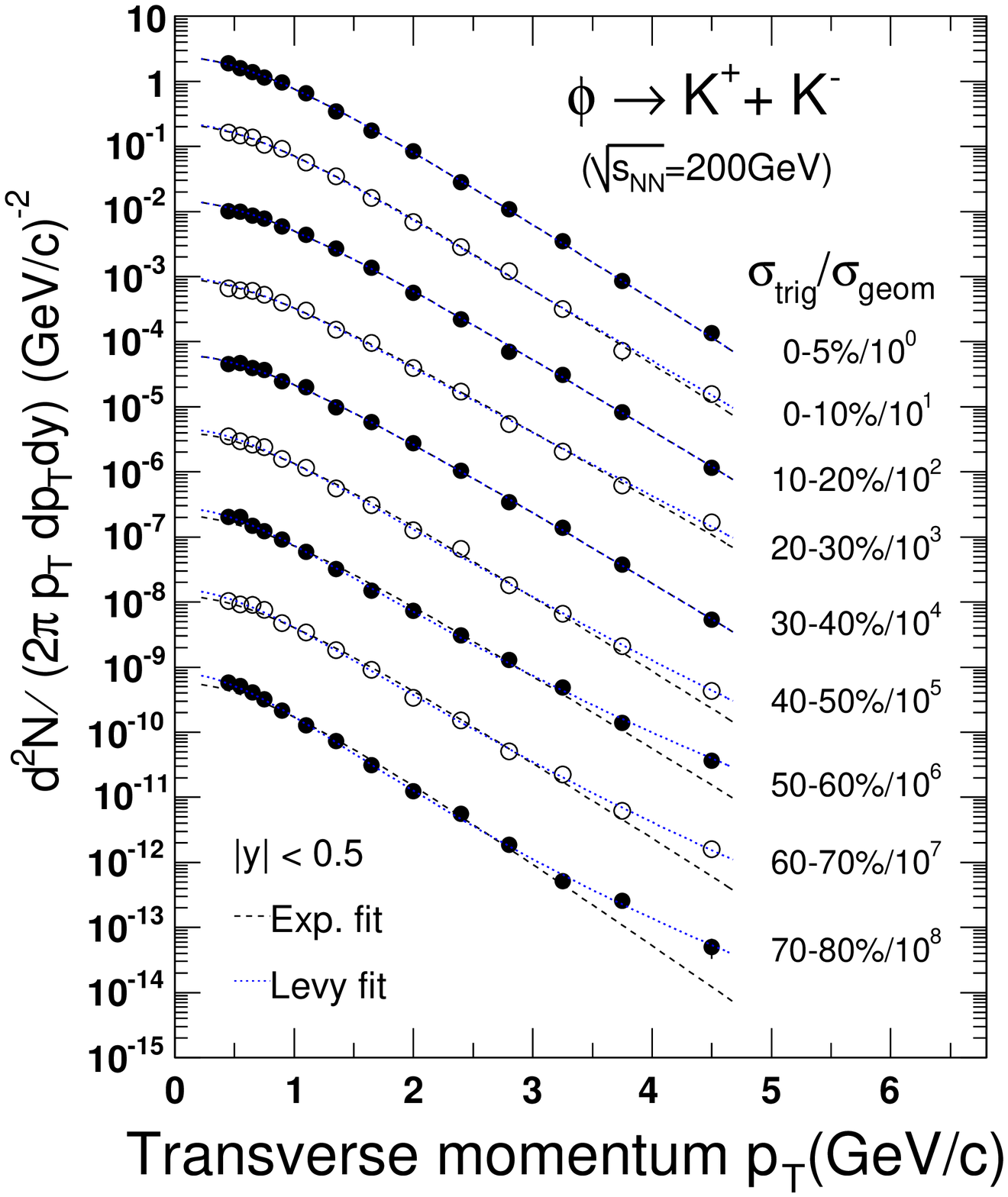}
  \includegraphics[scale=0.25]{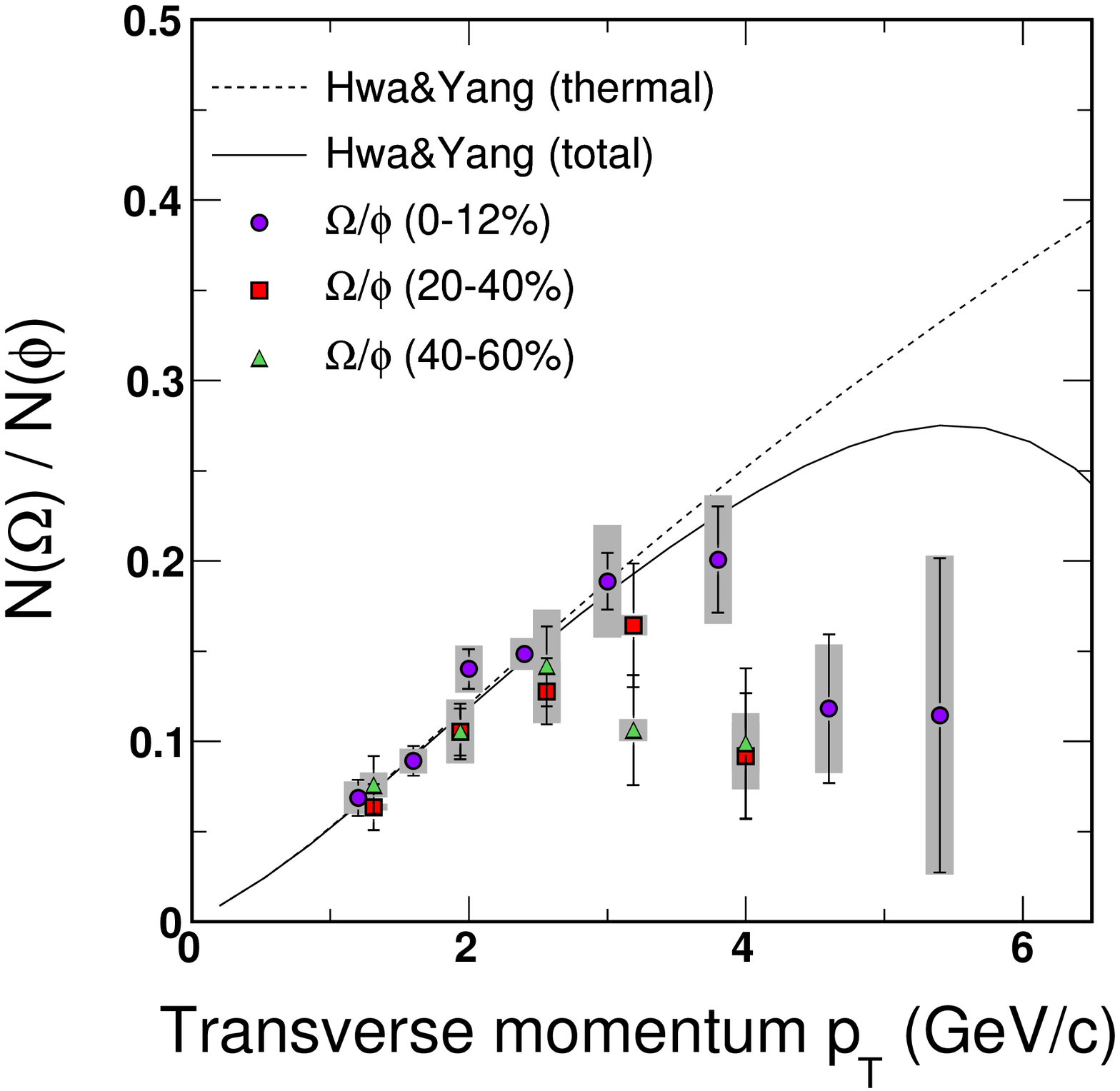}
\vspace{-0.55cm}
  \caption{\label{fig:Spectra}(color online) Left : $p_{T}$ distributions of $\phi$-mesons
    from Au+Au collisions at $\sqrt{s_{NN}}$ = 200 GeV. Right : The $N(\Omega)/N(\phi)$ ratio
  vs.\ $p_{T}$ for three centrality bins in $\sqrt{s_{NN}}$ = 200 GeV
  Au+Au collisions. The solid and dashed lines represent
  recombination model predictions for central collisions~\cite{RudyHwa} for total and thermal
  contributions, respectively.}
\end{figure}

Figure~\ref{fig:Spectra} (left) shows the $p_{T}$ distributions of
$\phi$-mesons as a function of centrality. 
Each $p_{T}$ spectrum in Fig.~\ref{fig:Spectra} has been fitted using
both an exponential function (dashed lines) in $m_T$ and a Levy function
(dotted lines) which has an exponential-like shape at low $p_{T}$ and
is power-law-like at higher $p_{T}$. While the central data are fitted
equally well by both functions the more peripheral spectra are better
fitted by the Levy function indicating less thermal contributions in
peripheral collisions. In Fig.~\ref{fig:Spectra} (right), the ratios of $N(\Omega)/N(\phi)$
vs.\ $p_{T}$ are presented as a function of centrality. 
Also shown in the figure are recombination model expectations for central
collisions~\cite{RudyHwa} based on $\phi$ and $\Omega$ production from
coalescence of thermal and shower $s$ quarks in the medium. The model describes
the trend of the data up to $p_{T}\sim4$ GeV/c but fails at higher
$p_{T}$. With decreasing centrality, the
observed $N(\Omega)/N(\phi)$ ratios seem to turn over at successively
lower values of $p_{T}$ indicating a smaller contribution from thermal
quark coalescence in more peripheral collisions. This is also
reflected in the smooth evolution of the spectra shapes from the
thermal-like exponential to power-law shapes shown in
Fig.~\ref{fig:Spectra}. These results indicates that the bulk of 
the $\phi$-mesons are made via coalescence 
of seemingly thermalized $s$ quarks in central Au+Au collisions.

\section{$\phi$-meson anisotropy - partonic collectivity}

\begin{figure}[!hbt]
  \center
  \includegraphics[width=0.35\textwidth,height=0.45\textheight]{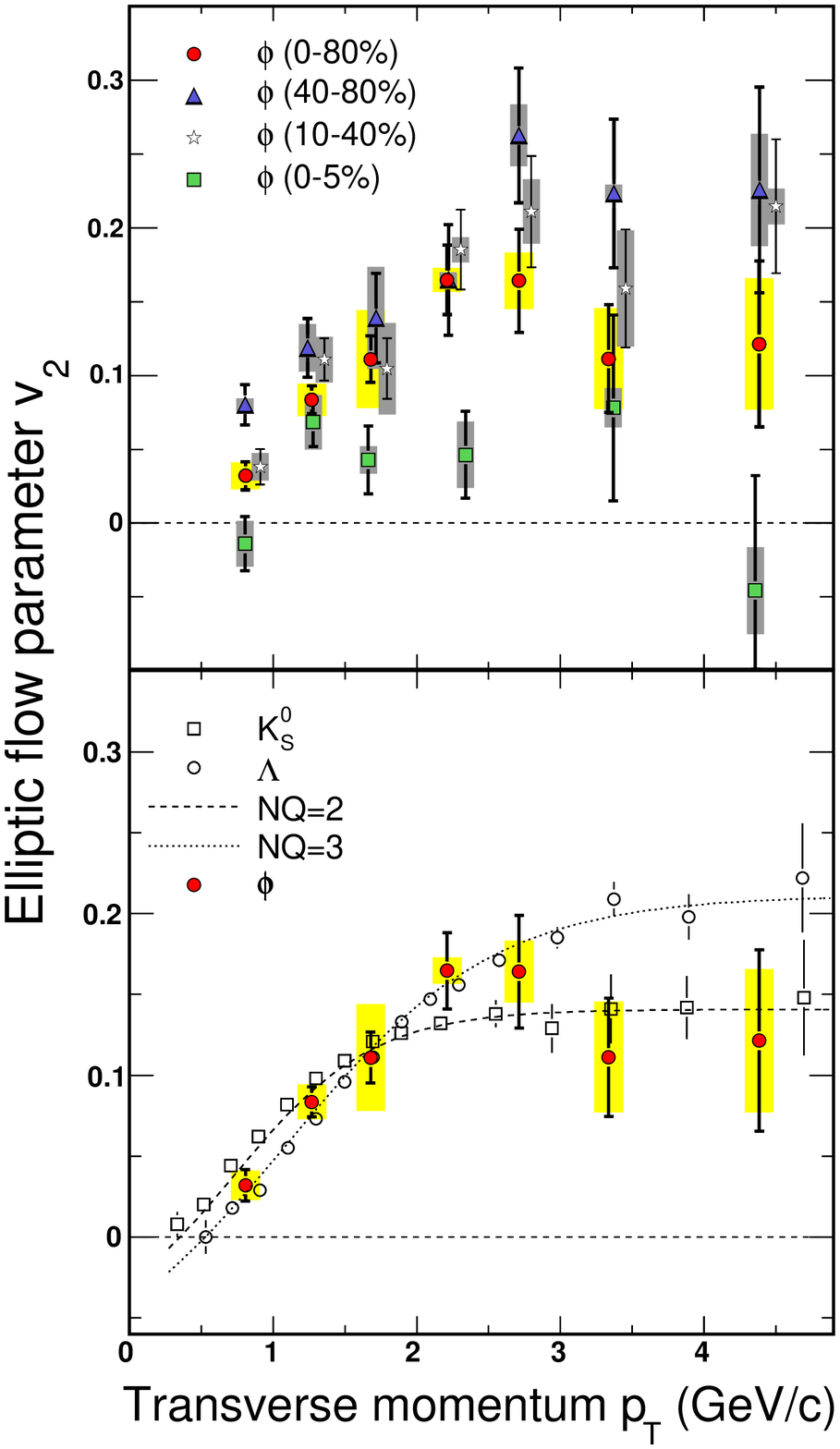}
  \includegraphics[scale=0.4]{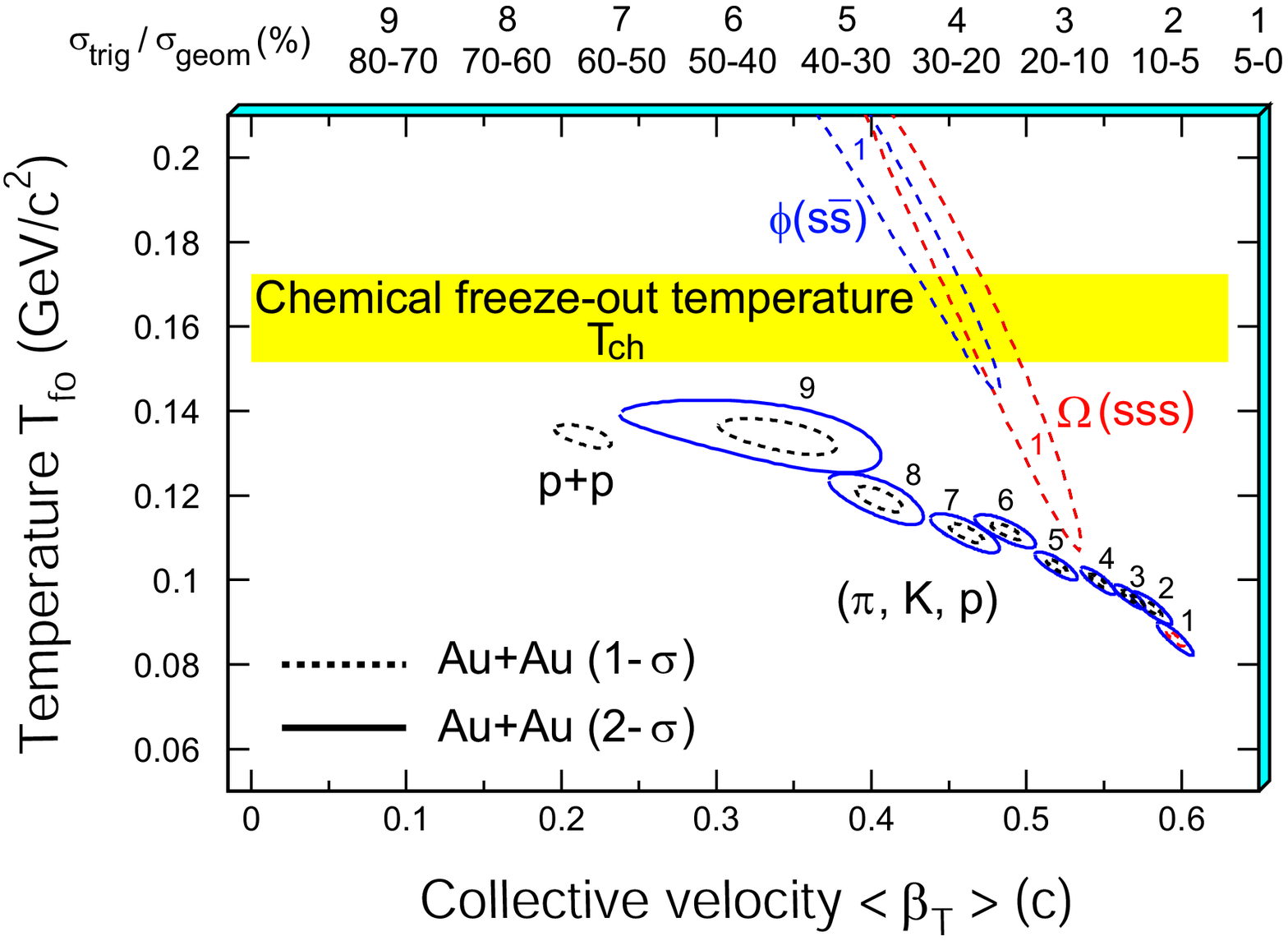}
  \vspace{-0.45cm}  
  \caption{\label{fig:V2}(color online) Left: Top panel: $v_{2}(p_{T})$, 
   for the $\phi$-meson as a function of centrality. 
    Bottom panel: Minimum bias $v_{2}(p_{T})$ for the
    $\phi$-meson compared to results for $\Lambda$ and $K^{0}_{S}$~\cite{star}.
    The dashed and dotted lines represent parameterizations
    of number-of-quark scaling, for NQ=2 and NQ=3
    respectively.
    Right : The $\chi^2$ contours, extracted from thermal +
radial flow fits, for produced
hadrons $\pi, K$ and $p$ and multi-strange hadrons $\phi$ and
$\Omega$. For $\phi$ and $\Omega$, only the most
central results are presented. Dashed and
solid lines are the 1-$\sigma$ and 2-$\sigma$ contours,
respectively.} 
\end{figure}

The top panel of Fig.~\ref{fig:V2} (left) shows the $v_{2}(p_{T})$ 
of the $\phi$-meson from Au+Au collisions for four centrality bins. 
The lower panel of Fig.~\ref{fig:V2} (left) shows the minimum bias (0-80\%) result compared to 
parameterizations for number-of-quark scaling for mesons (NQ=2) and baryons (NQ=3) 
whose free parameters have been fixed by fitting to the $\Lambda$ and $K^{0}_{S}$ results
simultaneously~\cite{star}.  In this case, for $p_{T} <$ 2 GeV/$c$, the
$\phi$ $v_{2}$ follows a mass-ordered hierarchy where the values of
$v_{2}$, within errors, fall between those of the heavier $\Lambda$
and lighter $K^{0}_{S}$. However, at
intermediate $p_{T}$, between 2-5 GeV/$c$, the $\phi$ $v_{2}$ appears
to follow the same trend as $K_S^0$. When we fit the $v_{2}$($p_T$) of
$\phi$-mesons with the quark number scaling ansatz~\cite{star}, the
resulting fit parameter NQ $=2.3\pm0.4$. The fact that the $\phi$
$v_{2}(p_{T})$ is the same as that of other mesons indicates that the
heavier $s$ quarks flow as strongly as the lighter $u$ and $d$
quarks. 

In order to show that $\phi$ $v_{2}$ reflects partonic collectivity
we have to demonstrate that (a) $\phi$-mesons are not formed through
kaon coalescence and (b) $\phi$-mesons do  not participate strongly in 
hadronic interactions and freezes out early from the system.
As previously discussed, $\phi$-mesons are not formed through
kaon coalescence~\cite{star}. In support for (b) we discuss 
the following observations from both experiments and theoretical models.
The general trend for $\bar{p}$, $K^-$ and $\pi^-$ is an increase in
$\langle p_T \rangle$ as a function of centrality, which is indicative
of an increased transverse radial flow velocity component to these
particles' momentum distributions (Fig.~\ref{fig6} top right)~\cite{star}. 
The $\phi$ $\langle p_T \rangle$, however, shows no significant centrality dependence. 
This indicates that the $\phi$ do not participate in the later hadronic rescatterings 
as that of $\bar{p}$, $K^-$ and $\pi^-$. This is expected if the $\phi$
decouples early on in the collision before transverse radial flow is
completely built up. If the $\phi$ hadronic scattering cross section
is much smaller than that of other particles, one would not expect the
$\phi$ $\langle p_T \rangle$ distribution to be appreciably affected
by any final state hadronic rescatterings. In contrast to these
observations, the RQMD predictions of $\langle p_T \rangle$ for kaon,
proton and $\phi$ all increase as functions of centrality
\cite{rqmd,star}.  Phenomenological analysis~\cite{seqgp} has suggested a
relatively small hadronic interaction cross section for $\phi$-mesons.
Further the data on coherent $\phi$ photoproduction shows that $\sigma_{\phi N}$
$\sim$ 10 mb~\cite{photo}. This is about a  factor of 3 times lower than $\sigma_{\rho N}$ and 
$\sigma_{\pi N}$; about a factor 4 times lower than $\sigma_{\Lambda N}$ and
$\sigma_{N N}$ and about a factor 2 times lower than $\sigma_{K N}$. 
A hydrodynamical (and hydro motivated blast wave) description of the $\phi$-meson 
spectra clearly indicates that the freeze-out temperature of $\phi$-mesons 
(Fig.~\ref{fig:V2} right, 160-190 MeV) at RHIC
energies is close to the critical temperature ($T_{C}$) for quark-hadron phase transition
predicted by Lattice QCD calculations~\cite{starwhitepapers}. 
All these indicate that $\phi$-mesons decouple
early from the system and do not participate strongly in
hadronic interactions. In addition discussion in previous section indicates that
$\phi$-mesons are produced from seemingly thermalized s-quarks. 
Therefore the $\phi$-meson $v_{2}$ results demonstrate partonic
collectivity.

\section{$\phi$-meson anisotropy - probe the QCD phase diagram}

In the previous sections we discussed the following features for $\phi$-mesons at RHIC:
(a) primodial contrinution close to 100\%, (b) decouples early from the system and
close to $T_{C}$, (c) at higher energies not formed from K$\bar{K}$ coalesence,
(d) enhancement reflects dense partonic medium formation, (e) formed by
coalesence of seemingly thermalized $s$$\bar{s}$ quarks and  (f) substantial collectivity 
which at intermediate $p_{T}$ exhibits NQ scaling. All these indicate the large $\phi$ $v_{2}$ 
is developed due to partonic interactions in the medium.
Based on these observations we propose that, absence and reduction of $\phi$ $v_{2}$ compared
to other hadrons and absence of NQ scaling for $\phi$ $v_{2}$ (note $\phi$-mesons so far is 
the only hadron that can differentiate between mass and NQ effects upto RHIC energies) 
indicate during the evolution the system remains in the hadronic phase. 
Since the number of quark scaling requires the deconfinement to take place. 
Therefore one does not expect such a scaling in a pure hadronic system.
More supporting arguments that 
substantial $\phi$ $v_{2}$ can only arise from partonic interactions comes from
the transport model calculations.  Both RQMD and UrQMD calculations failed to
explain the substantial $v_{2}$ for charged hadrons measured as a function
of collision centrality in Au+Au collisions at 200 GeV in RHIC~\cite{rqmd}. Thereby leaving
for a substantial scope for partonic interactions to generate the remaining $v_{2}$.

\begin{figure}[!hbt]
\vspace{-0.45cm}
  \center
  \includegraphics[scale=0.6]{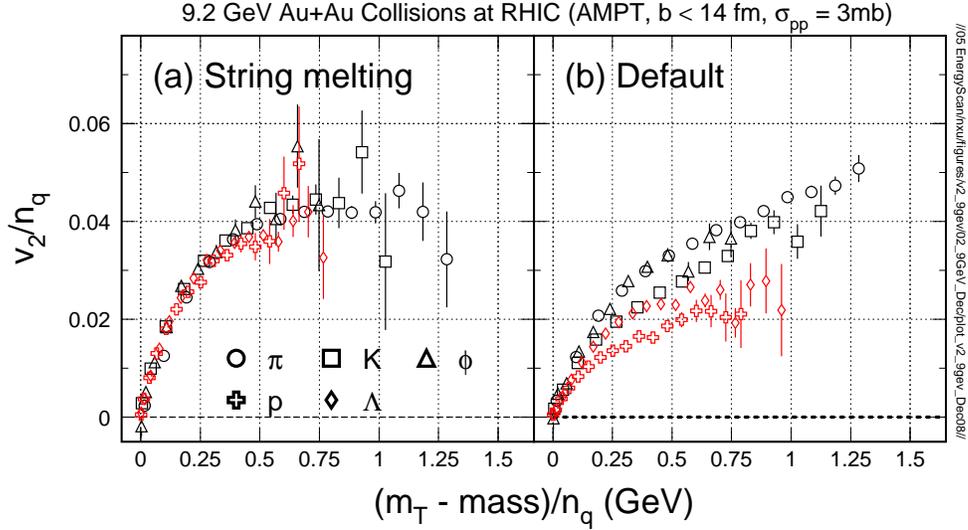}
\vspace{-0.55cm}
  \caption{\label{ampt}(color online) Elliptic flow ($v_{2}$) of $\phi$, $\pi$, $K$, $p$, $\Lambda$ normalized by NQ
   from Au+Au collisions at $\sqrt{s_{NN}}$ = 9.2 GeV at midrapidity for impact parameter less 
   than 14 fm. The results from AMPT model~\cite{ampt} are plotted as a function of 
   $\langle$ $m_{T}$ $\rangle$ - $m$ normalized by NQ.
  Left-plot : String melting scenario where partonic interactions were mimicked via 
  breaking a string into quarks and the hadronization is achieved by quark 
  recombination. Right-plot: Default case where only hadronic interactions were 
  employed. This is the case most close to the low energy collisions below the 
 possible phase diagram.}
\end{figure}

In Fig.~\ref{ampt} we show $\phi$ $v_{2}$ from A Multi Phase Tramsport Model (AMPT)~\cite{ampt} 
as a function of difference in average transverse mass 
($\langle$ $m_{T}$ $\rangle$) and hadron mass ($m$) both normalized by the number of 
constituent quarks (along with other hadrons: $\pi$, $K$, $p$, $\Lambda$) 
for two cases (a) right plot shows the results with default setting and (b) left plot shows 
the results with string melting scenario, which incorporates partonic
coalescence mechanism. All results are for Au+Au collisions at $\sqrt{s_{NN}}$ = 9.2 GeV, for
-1.0 $<$ $\eta$ $<$ 1.0 and impact parameter less than 14 fm. Clearly one observes
breaking of NQ scaling for the results in the default case and excellent NQ scaling for
the string melting case. This indicates the $\phi$ $v_{2}$ can be used to probe if the 
relevant degrees of freedom for the medium formed in heavy ion collisions is partonic
or hadronic. One may question the high values of $\phi$ $v_{2}$ in this model for default case.
This could be because of other mechanism of $\phi$ production incorporated in the model, like
from example K$\bar{K}$ coalescence. However results from RHIC data has indicated that such
mechanisms are not dominant. Those processes if taken out could possibly lead to small $\phi$ $v_{2}$
in the default case. In short, two predictions from the model analysis: without the formation 
of partonic matter we expect to observe (i) the absence of the $v_2$ NQ scaling for 
all hadrons; (ii) small or zero $v_2$ for the $\phi$-mesons.

\section{Summary}

Several interesting features of $\phi$-mesons are observed to provide important information
about the medium formed in heavy ion collisions. These are summarized in Table~\ref{table1}.
$\phi$-meson production in Au+Au and Cu+Cu collisions relative to $p$+$p$ collisions has 
shown that the strangeness enhancement in heavy-ion collisions at RHIC
is due to formation of a 
dense partonic medium and not due to Canonical suppression effects. Results on $\phi$-meson 
elliptic flow provides a clear evidence that the collectivity observed in heavy-ion 
collisions at RHIC is developed at the partonic stage. 
The results from the $N(\Omega)/N(\phi)$ ratios and comparison to model calculations indicate
the $\phi$-mesons are produced via coalescence of seemingly thermalized $s$-quarks. 
Based on the several interesting observation connected to $\phi$-meson production it is
possible to use the observation of large collectivity and number of constituent quark scaling 
of  $\phi$-meson elliptic flow as an indication that matter went through partonic phase. Hence
$\phi$ $v_{2}$ can be used as a probe for locating the QCD phase boundary. This can be 
explored in future RHIC beam energy scan program and at the compressed baryonic matter 
experiment at GSI.

{\it Acknowledgments:}
Authors would like to thank F. Becattini, J.H. Chen, J. Cleymanns, N. Herrmann, H. Huang, 
F. Liu, G. L. Ma, H. Oeschler, J. Rafelski, J. Takahasi, K.J. Wu and  Z. Xu for discussions 
on the topic of this paper at SQM2008.

\end{document}